\documentclass[prd,nofootinbib,reprint,twocolumn,superscriptaddress]{revtex4-1}
\pdfoutput=1 
\usepackage{bm, color} 
\usepackage{amssymb,amsfonts,slashed,amsthm,amsmath, graphicx, soul, empheq, cancel, hyperref}
\usepackage[caption=false]{subfig}

\newcommand{\bea}{\begin{array}}
\newcommand{\eea}{\end{array}}
\newcommand{\beq}{\begin{eqnarray}}
\newcommand{\eeq}{\end{eqnarray}}

\newcommand{\MeV}{\  {\rm MeV} }

\def\SEC#1{Sec.~\ref{#1}}

\def\FIG#1{Fig.~\ref{#1}}

\numberwithin{equation}{section}

\definecolor{orange}{RGB}{255,100,0}
\definecolor{rosepink}{RGB}{248,100,100}

\begin{document}

\title{
Cosmological Roles of Dark Photons in Axion-induced Electroweak Baryogenesis
}

\author{Kwang Sik Jeong}
\email{ksjeong@pusan.ac.kr}
\affiliation{Department of Physics, Pusan National University, Busan 46241, Korea}

\author{Ju Hyeong Kang} 
\email{kas6306@pusan.ac.kr}
\affiliation{Department of Physics, Pusan National University, Busan 46241, Korea}

\author{Shota Nakagawa}
\email{shota.nakagawa@sjtu.edu.cn}
\affiliation{Tsung-Dao Lee Institute, Shanghai Jiao Tong University, \\
No.~1 Lisuo Road, Pudong New Area, Shanghai, 201210, China}
\affiliation{School of Physics and Astronomy, Shanghai Jiao Tong University, \\
800 Dongchuan Road, Shanghai 200240, China}

\begin{abstract}

By coupling to both the Higgs and electroweak gauge sectors, 
an axion can generate the matter-antimatter asymmetry in the universe via
electroweak baryogenesis
when the axion decay constant lies within the range of approximately $10^5$ and $10^7$~GeV, 
corresponding to axion masses between the MeV and GeV scales. 
In this work,  we explore the intriguing  possibility that the axion interacts with a dark sector, 
particularly with dark photons through anomalous couplings. 
Notably, 
axion-coupled dark photons can play multiple roles,  including 
$(i)$ suppressing the branching ratio of axion decay to Standard Model (SM) particles,  
which would otherwise conflict with the constraints from supernovae explosions, Big Bang nucleosynthesis, and neutron star merger events, 
$(ii)$
serving as a candidate for cold dark matter if they are massive and stable,
and $(iii)$  contributing to dark radiation if they are ultralight. 
The axion decouples from the SM thermal bath when it becomes non-relativistic, 
facilitating the production of dark matter dark photons through the freeze-in mechanism,  
while dark radiation dark photons are thermally generated prior to the electroweak phase transition.
\end{abstract}

\maketitle

\section{Introduction
\label{introduction}}

Axions, also referred to as axion-like particles, 
can play a crucial role in refining the Standard Model (SM) of particle physics. 
They enhance the naturalness of the SM by addressing issues such as the strong CP problem~\cite{Peccei:1977hh,Weinberg:1977ma,Wilczek:1977pj} and the gauge hierarchy problem~\cite{Graham:2015cka}. 
Additionally, they are compelling candidates for unknown degrees of freedom, including dark
matter~\cite{Preskill:1982cy,Abbott:1982af,Dine:1982ah,Arias:2012az,Chadha-Day:2021szb,Adams:2022pbo} and inflaton~\cite{Freese:1990rb}. 
Moreover, they offer a plausible explanation for the observed 
excess of matter over antimatter in the universe~\cite{Takahashi:2003db,Servant:2014bla,Jeong:2018ucz,Jeong:2018jqe,Co:2019wyp,Domcke:2020kcp,Chakraborty:2021fkp,Im:2021xoy,Harigaya:2023bmp,Chun:2023eqc}.
Notably, these particles can be explored through astrophysical and cosmological observations, as well as laboratory experiments~\cite{Raffelt:2006cw,Marsh:2015xka,Graham:2015ouw,Irastorza:2018dyq,Bauer:2017ris,Sikivie:2020zpn,Caputo:2024oqc}.

In this paper, we focus on an axion that feebly couples to the Higgs sector, as this coupling enables the implementation of electroweak baryogenesis while remaining consistent with experimental constraints on electric dipole moments (EDM).
Through its interaction with the Higgs sector, the axion can facilitate a strong first-order electroweak phase transition~\cite{Jeong:2018ucz}. 
Furthermore, if the axion couples to the electroweak gauge sector,  substantial $CP$ violation is achieved during its evolution.
In the presence of these couplings,
the observed baryon asymmetry can be 
adequately explained if the axion decay constant lies within the range of approximately $10^5$ to $10^7$~GeV,
corresponding to axion masses ranging from the MeV to GeV scales~\cite{Jeong:2018jqe,Harigaya:2023bmp}.

The axion responsible for baryogenesis couples to both the Higgs and electroweak gauge sectors, 
with the coupling strength suppressed by the axion decay constant. 
As a result, it is subject to various cosmological, astrophysical, and experimental constraints.
In particular,  since its mass is approximately the electroweak scale squared divided by the decay constant,  the axion anomalous coupling to photons is severely constrained by standard Big Bang nucleosynthesis (BBN)~\cite{Cadamuro:2011fd}, supernovae (SN) explosions~\cite{Jaeckel:2017tud,Caputo:2021rux,Caputo:2022mah,Hoof:2022xbe,Muller:2023vjm,Diamond:2023scc}, 
and neutron star merger events~\cite{Diamond:2023cto,Dev:2023hax}, 
especially within the broad range of axion decay constants necessary for baryogenesis.
However,  the allowed parameter space can be significantly expanded if the axion predominantly 
decays to light dark particles rather than SM particles.

The framework of axion-induced baryogenesis currently lacks a candidate for dark matter. 
In this work, we investigate the intriguing possibility that the axion also plays a significant role in dark matter production, under the assumption that its interactions are governed by 
the associated global U$(1)$ symmetry. 
We propose the inclusion of dark photons,  which couple to the axion, 
as a candidate for dark matter. 
Notably,  
axion-coupled dark photons can serve multiple roles depending on their mass, 
such as constituting the dark matter,
relaxing the astrophysical constraints on the axion coupling to the electroweak gauge 
sector, and contributing dark radiation 
with various cosmological implications.

This paper is organized as follows. 
In Section \ref{sec:axion}, we begin with a brief review of the mechanism by which the axion induces a strong first-order electroweak phase transition. 
We subsequently examine the current constraints on axion-Higgs mixing,  through which the axion interacts with other SM particles,  as well as the constraints on the axion coupling to the electroweak gauge sector. 
Additionally,  we delve into the detailed process  of axion thermalization in the early universe.
In Section \ref{sec:dark-photon},  we introduce dark photons that interact with the axion
through an anomalous coupling.
We analyze three distinct types of dark photons, 
each serving different cosmological roles. 
Finally, Section \ref{sec:conclusion} presents our conclusions.

\section{Axion-induced electroweak baryogenesis
\label{sec:axion}}

\subsection{Bayogenesis 
\label{sec:baryogenesis}}

In this subsection, we provide a brief review of axion-induced electroweak baryogenesis~\cite{Jeong:2018jqe}.
The  axion, denoted by $\phi$, interacts with other fields through couplings suppressed 
by its decay constant $f_\phi$. 
These interactions are governed by a global U$(1)_\phi$ symmetry,  under which the axion transforms as  $\phi \to \phi + {\rm constant}$,  
a symmetry assumed to be preserved at the perturbative level. 
We extend the Higgs sector to include the axion through the potential 
\begin{equation}
    V = \frac{\mu^2_{\rm eff}(\phi)}{2}h^2
    + \frac{\lambda}{4}h^4
    + V_0(\phi),
\end{equation}
where $h$ is the neutral Higgs field.
The periodicity, $\phi\to \phi + 2\pi f_\phi$, 
necessitates that the potential be a function of $\phi/f_\phi$.
As a concrete model,  we consider
\beq
\mu^2_{\rm eff} &=& \mu^2 - M^2 \cos\left(
\frac{\phi}{f_\phi} + \beta  \right),
\nonumber \\
V_0 &=&
    - \Lambda^4 \cos
    \left( \frac{\phi}{f_\phi}\right)
    + {\rm constant},
\eeq
where $\mu$,  $M$,  and $\Lambda$ are 
parameters of the order of the electroweak scale.
A natural UV completion of the axion coupling to the Higgs sector 
involves the introduction of a new confining gauge group, 
whose fermions are charged under  U$(1)_\phi$
and couple to the Higgs field via Yukawa interactions~\cite{Graham:2015cka,Jeong:2018jqe}.

The axion varies the effective Higgs mass squared 
and can therefore play a crucial role in the electroweak phase transition if $M$ is larger than $\mu$. 
For large values of $f_\phi$, the axion couples feebly to the Higgs sector, distinguishing this scenario from other extensions of the Higgs sector.
Despite its weak coupling, it has been observed that the axion can induce a first-order electroweak phase transition over a 
wide parameter space~\cite{Jeong:2018jqe,Jeong:2019ple,Harigaya:2023bmp}.
Interestingly, 
some of the parameter space may be accessible to current and upcoming experimental probes because the axion mixes with the Higgs boson after electroweak symmetry breaking.

Let us briefly examine the properties of the axion for $f_\phi$ much larger than the electroweak scale.
The axion mass is given by 
\begin{equation}
\label{axion-mass}
    m_\phi = \kappa_1 \frac{m^2_h}{f_\phi},
\end{equation}
and the axion-Higgs mixing angle is determined by
\begin{equation}
\label{axion-Higgs-mixing}
    \sin\theta_{\rm mix} = 
    \kappa_2
    \frac{m_h}{f_\phi},
\end{equation}
with $m_h\simeq 125$~GeV being the Higgs boson mass.
Here the constants $\kappa_1$ and $\kappa_2$
are determined by the model parameters, \footnote{
To examine the vacuum structure of the potential, it is useful to introduce the dimensionless parameters:
\begin{equation}
\epsilon \equiv \frac{\sqrt{2\lambda} \Lambda^2 }{M^2},
\quad
r \equiv \frac{\sqrt2 \Lambda^2 }{\sqrt\lambda v^2},
\nonumber 
\end{equation}
where $v=246$~GeV is the Higgs vacuum expectation value. 
A first-order electroweak phase transition requires $\epsilon <1$~\cite{Jeong:2019ple}.
The minimization conditions of the potential determine the values of
$\kappa_1$ and $\kappa_2$ as  
\beq
\kappa_1 &\simeq& 
\sqrt{
\frac{
r(1+2r\epsilon \cos\beta+r^2\epsilon^2)^{3/2} - r^4\epsilon \sin^2\beta
}
{8\lambda \epsilon
(1+ 2r\epsilon \cos\beta + r^2 \epsilon^2)
}
},
\nonumber \\
\kappa_2 &\simeq& 
\frac{r^2  \sin\beta}{\sqrt{
8\lambda(1+ 2r\epsilon \cos\beta + r^2 \epsilon^2)
}},
\nonumber 
\eeq
for $f_\phi$ much above the electroweak scale.
\label{foot:param}
} 
and a strong first-order electroweak phase transition is achieved within a broad parameter 
space where $\kappa_1$ and $\kappa_2$ are of order unity or less.
The mixing results in interactions between the axion and SM particles beyond the 
Higgs boson.

At high temperatures, the universe remains in the symmetric phase with 
$(h,\phi) = (0,0)$ due to 
large thermal corrections to the Higgs scalar potential. 
As the universe cools, the potential develops an electroweak minimum at 
$\phi \neq 0$, 
separated from the symmetric phase by a potential barrier. 
As the temperature decreases further and the electroweak minimum deepens, 
the symmetric false vacuum decays through the nucleation of critical bubbles of the broken phase,
provided the barrier persists long enough. 
The bubble nucleation rate is given by $\Gamma \propto T^4 e^{-S_3/T}$, 
where $S_3$ represents the Euclidean action of an O$(3)$ symmetric critical bubble. 
For $f_\phi$ much above the TeV scale, tunneling predominantly occurs along the axion direction, with $S_3/T$ approximately proportional to 
$(T - T_2)^2 f^3_\phi$, where
$T_2$ denotes the temperature at which the barrier disappears. 
Consequently, a first-order phase transition is achieved across
a broad parameter space for large $f_\phi$.
It is noteworthy that 
the bubble wall thickness scales approximately with $f_\phi/\Lambda^2$,  
and the phase transition proceeds relatively smoothly with the nucleation of bubbles.
The phase transition is thus nearly adiabatic during baryogenesis, 
and diffusion through the bubble wall is inefficient.

The axion, when coupled to the Higgs sector, can facilitate 
a strong first-order electroweak phase transition. 
Moreover, if it couples to the electroweak gauge sector via 
\begin{equation}
\label{axion-SU2}
    \frac{\phi}{f_\phi} W \tilde{W},
\end{equation}
the axion can naturally induce local spontaneous 
baryogenesis~\cite{Cohen:1987vi,DeSimone:2016ofp}, 
as the time derivative of the axion field acts as a chemical potential for 
the Chern-Simons number at a given spatial point.
In the presence of the above coupling,
the baryon number is generated through the electroweak anomaly, governed by
\begin{equation}
    \frac{d n_B}{dt} \approx\frac{N_g}{2} \frac{\Gamma_{\rm sph}}{T}
    \frac{d}{dt}\frac{\phi}{f_\phi} - \Gamma_B n_B,
\end{equation}
where $N_g=3$ is the number of generations,
 $\Gamma_{\rm sph}$ represents the electroweak sphaleron transition rate per unit volume,
and $\Gamma_B = (13N_g/4) \Gamma_{\rm sph}/T^3$ is
the sphaleron-induced washout rate~\cite{Bochkarev:1987wf,Cline:2000nw}.

The axion-Higgs mixing causes the axion to dissipate energy into the background plasma 
through interactions with SM particles.
As a result, the axion undergoes underdamped oscillations within 
the nucleated bubbles while generating baryon asymmetry.
More precisely, the axion evolution can be divided into two stages: 
an initial phase of falling toward the potential minimum, 
followed by oscillations around this minimum. 
Baryon asymmetry is efficiently generated during the initial fall, 
but it is washed out in the subsequent oscillatory phase 
as the sign of $d\phi/dt$ changes, leading to the partial cancellation between 
baryon and anti-baryon numbers. 
The washout is effective only in the region of $\phi$ where the Higgs background field value
\begin{equation}
  \langle h \rangle_T \simeq \sqrt{-\frac{\mu^2_{\rm eff}(\phi)+c_h T^2}{\lambda}}
\end{equation}
is smaller about $0.5T$, as otherwise the sphaleron rate becomes exponentially 
suppressed~\cite{DOnofrio:2014rug}. 
Here $c_h$ is a positive constant, determined by SM couplings, 
representing the thermal correction to the Higgs mass squared.
Thus, strong washout can be avoided 
if the thermal friction is sufficiently large to quickly dampen the axion oscillation amplitude following the initial fall.

As demonstrated in Ref.~\cite{Jeong:2018jqe}, the  baryon-to-entropy ratio
generated through axion-induced spontaneous baryogenesis can be expressed as
\begin{equation}
    \frac{\eta_B}{\eta_B|_{\rm obs}}
   = \kappa_B \left( \frac{T_n}{60~{\rm GeV}} \right)^2,
\end{equation}
where $\eta_B|_{\rm obs}\simeq 0.92\times 10^{-10}$ is the observed baryon asymmetry.
The parameter $\kappa_B$, which depends on the model parameters and is of order unity 
or smaller, 
quantifies how much the baryon asymmetry is washed out during axion oscillations
after the initial fall.
The bubble nucleation temperature  $T_n$, at which the nucleation rate becomes comparable to the Hubble expansion rate, is approximated as 
\begin{equation}
    T_n \approx
    \sqrt{   1
    - 2\frac{M^2}{m^2_h}
    \left(  \cos\left(\frac{\phi_0}{f_\phi} +\beta \right) - \cos\beta
    \right)  } \, T^{\rm SM}_c,
\end{equation}
with $T^{\rm SM}_c\simeq 160$~GeV,
the critical temperature for the electroweak phase transition in the SM. 
In the model under consideration, $T_n$ is close to the barrier disappearance temperature $T_2$.

The thermal friction during axion evolution is predominantly caused by its coupling to top quarks, which arises through mixing with the 
Higgs boson~\cite{Wang:1999mb,Mukaida:2012qn}.
For $f_\phi\lesssim 10^7$~GeV, this thermal friction is sufficiently strong, 
ensuring that 
$\kappa_B$ remains of order unity across the parameter space where a strong first-order electroweak phase transition can occur.
In this regime, $\kappa_B$ is relatively insensitive to the precise value of $f_\phi$.
This indicates that the observed baryon asymmetry can be successfully generated via spontaneous baryogenesis.
However, for larger values of $f_\phi$, the washout effect becomes increasingly significant, resulting in a substantial suppression of $\kappa_B$
to values well below unity.

\subsection{Constraints on axion
\label{sec:constraint}}
 
Due to its coupling to both the Higgs and electroweak sectors,  
the axion is subject to various experimental, astrophysical, and cosmological constraints. 
Since the earlier studies~\cite{Jeong:2018jqe,Jeong:2018ucz}, 
recent updates to existing constraints necessitate a re-evaluation of the allowed parameter space for our scenario.  
These constraints impose restrictions on axion-Higgs mixing, 
$\phi h$,  with the mixing angle given by Eq.~(\ref{axion-Higgs-mixing}),
as well as on the axion-photon interaction, 
$\phi F_{\mu\nu}\tilde F^{\mu\nu}$,  with a
coupling given by
\begin{equation}
    g_{\phi\gamma} = \frac{c_\gamma \alpha}{2\pi f_\phi},
\end{equation}
where $\alpha=e^2/(4\pi)$ is the fine-structure constant,
and $c_\gamma$ denotes
the anomaly coefficient, a rational number typically of order unity.\footnote{ 
In Eq.~(\ref{axion-SU2}),  the anomaly coefficient for the coupling to the SU$(2)_L$ gauge bosons has been set to unity for simplicity.
However, more generally, it can be a rational number of order unity. 
Additionally, the axion-photon coupling, $g_{\phi\gamma}$,  receives a further contribution 
if the axion also couples to the U$(1)_Y$ gauge bosons. 
}
The allowed range of these couplings depends on the axion mass.
Note that the Higgs boson has a loop-induced coupling to photons,
and thus axion-Higgs mixing induces the interaction, 
$\phi F_{\mu\nu}F^{\mu\nu}$.
However,  this effect remains subdominant unless 
$c_\gamma$ is much smaller than order unity.

\begin{figure}[t!]
\centering
\includegraphics[width=8.5cm]{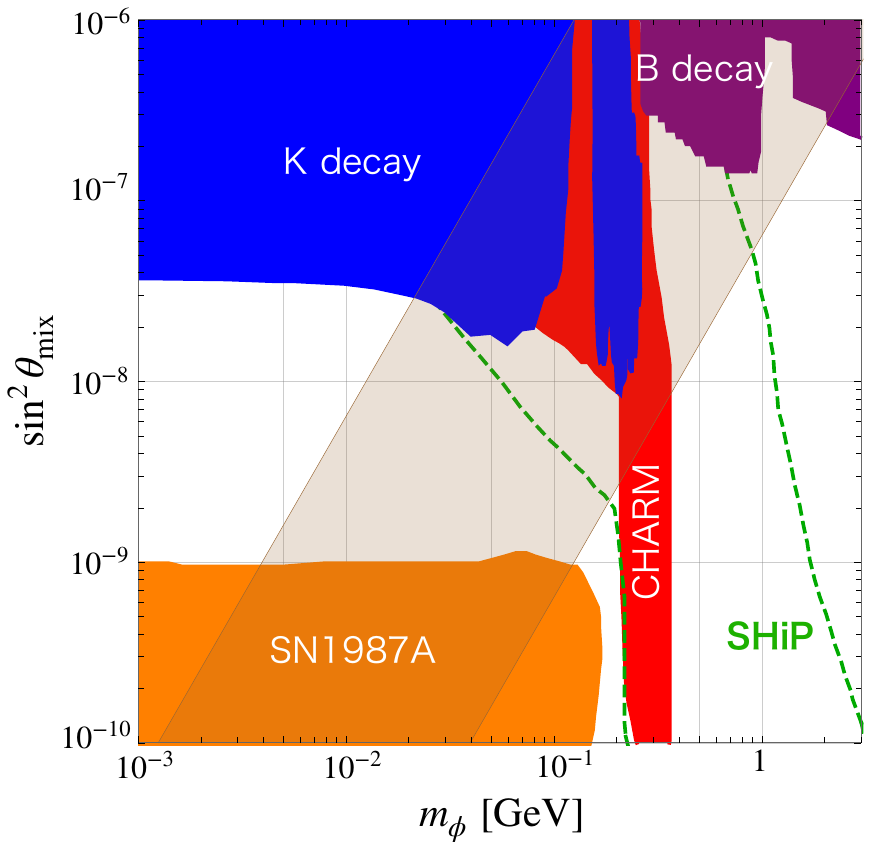}
\caption{
Constraints on axion-Higgs mixing as a function of the axion mass. 
The light brown band represents the relationship between the axion-Higgs mixing angle  
and the axion mass,  given by
$\sin^2\theta_{\rm mix} = \kappa_{\rm mix}  (m_\phi/m_h)^2$,  
for values of $\kappa_{\rm mix}$ ranging from $10^{-3}$ to $1$.  
The colored shaded regions are excluded by rare meson decays (K-meson for blue and B-meson for purple), beam-dump searches for axion-like particles (red),  and supernova cooling (orange),  respectively. 
The area enclosed by the green dashed lines indicates a region that will be explored by SHiP \cite{Alekhin:2015byh}. 
}
\label{fig:bound-on-mixing}
\end{figure}

First,  we investigate the constraints on axion-Higgs mixing. 
The mixing is limited by the radiative contribution of the axion to 
the electron EDM,  which is expressed as~\cite{Choi:2016luu}
\begin{equation}
    d_e \simeq 
    \frac{e^3}{32\pi^4}\frac{m_e}{v}
    \frac{\sin\theta_{\rm mix}}{f_\phi}
    \ln\left(\frac{m_h}{m_\phi}\right).
\end{equation}
This contribution remains below the ACME~II and 
JILA limits~\cite{ACME:2018yjb,Roussy:2022cmp} 
for $f_\phi$ larger than about $10$~TeV. 
Notably,  many other electroweak baryogenesis scenarios face stringent EDM constraints, 
as achieving a strong first-order phase transition typically necessitates fields 
that couple strongly to the Higgs sector.
Furthermore, the axion-Higgs mixing 
is also constrained by experimental bounds  on rare meson decays \cite{E787:2004ovg,BNL-E949:2009dza,NA62:2021zjw,Belle:2009zue,LHCb:2012juf,LHCb:2015nkv} and beam-dump searches \cite{CHARM:1985anb} for axion-like particles.
Additionally,  
if the axion mass is below the SN core temperature,  axions can be produced abundantly in a core-collapse SN via interactions with nuclear matter. 
To prevent excessive SN cooling,  the axion-Higgs mixing should either be sufficiently 
small to suppress axion production or sufficiently large to trap axions 
within the SN core~\cite{Raffelt:1987yt,Burrows:1990pk,Carenza:2019pxu}.  
In our model, since successful baryogenesis requires a decay constant below $10^7$~GeV,  
the axion should reside in the trapping regime.  
A comprehensive summary of these constraints can be found in Refs.~\cite{Choi:2016luu,Flacke:2016szy,Winkler:2018qyg}.
The bound from NA62~\cite{NA62:2021zjw} has been incorporated, 
updating the previous analyses~\cite{Jeong:2018jqe,Jeong:2018ucz}.
These combined constraints impose
\begin{equation}
    10^{-9} \lesssim \sin^2\theta_{\rm mix} \lesssim 3\times 10^{-7},
\end{equation}
if the axion mass is in the range of approximately $1$~MeV and 
$2m_\mu$, with $m_\mu$ being the muon mass, and
\begin{equation}
     \sin \theta_{\rm mix}
    \lesssim \frac{0.8 \times 10^{-3}}{\sqrt{{\rm Br}(\phi\to \mu^+\mu^-)}},
\end{equation}
if the axion mass lies between approximately $300$~MeV and $5$~GeV.
Here ${\rm Br}(\phi\to \mu^+\mu^-)$ denotes the branching ratio for the axion decay into a muon pair.
Fig.~\ref{fig:bound-on-mixing} provides a summary of the constraints on axion-Higgs mixing under the assumption that the axion decays exclusively into SM particles. 
The light brown band illustrates the relation,
$\sin^2\theta_{\rm mix} = \kappa_{\rm mix} (m_\phi/m_h)^2$,
where $\kappa_{\rm mix}=(\kappa_2/\kappa_1)^2$ is expressed in terms of the model 
parameters as
\begin{equation}
\kappa_{\rm mix}
\simeq 
\frac{r^3 \epsilon \sin^2\beta}
{(1 +2 r \epsilon \cos\beta + r^2 \epsilon^2 )^{3/2} 
-r^3 \epsilon \sin^2\beta
},
\end{equation}
which ranges between approximately $10^{-3}$ and $1$
for  $\beta$ values on the order of $0.1$ to $1$,  as naturally expected, 
and for $r$ and $\epsilon$ values (defined in footnote\ref{foot:param}) on the order of unity or smaller, 
as required for a strong first-order electroweak phase transition.

\begin{figure}[t!]
\centering
\includegraphics[width=8.5cm]{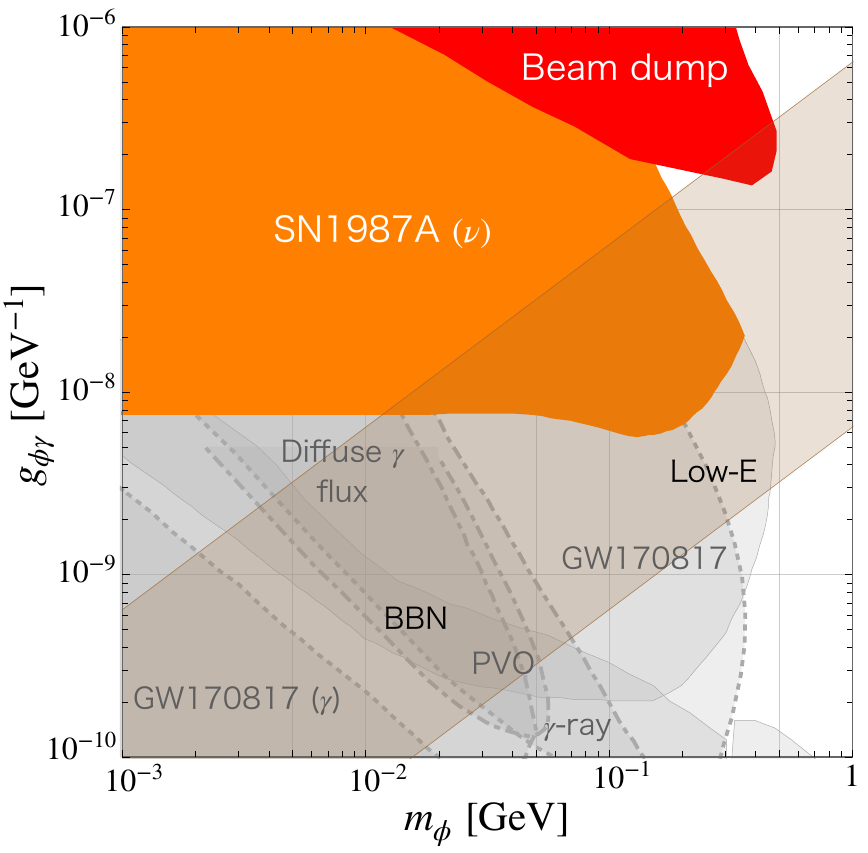}
\caption{
Constraints on the axion-photon coupling as a function of the axion mass. 
The light brown band illustrates the relationship between the axion-photon coupling 
and the axion mass,  given by 
$g_{\phi\gamma} = \kappa_{\phi\gamma}  m_\phi/m^2_h$, 
for values of $\kappa_{\phi\gamma}$ in the range of $10^{-4}$ and $10^{-2}$. 
The colored shaded regions indicate  
exclusions from beam-dump searches for axion-like particles (red) and supernova cooling 
constraints (orange),  respectively. 
The gray shaded areas, bounded by various gray lines corresponding to different constraints, indicate regions in tension with BBN, 
supernova observations, and neutron star merger events. 
However, these tensions can be mitigated if axion-Higgs mixing or the presence of an axion-coupled dark sector results in sufficiently rapid axion decay
and suppression of the branching ratio of axion decay into
SM particles.
}
\label{fig:bound-on-axion-to-photons}
\end{figure}

We now turn to examining the restrictions on the axion coupling to photons. 
Given the relation
\begin{equation}
\label{axion-mass-g}
    m_\phi \sim 10\,{\rm MeV}
    \left(\frac{g_{\phi\gamma}}{10^{-9}\,{\rm GeV}^{-1}}\right),
\end{equation} 
significant constraints arise from SN explosions,  neutron star merger events, 
and the predictions of standard BBN,  
particularly for the values of $f_\phi$
required for the successful  baryogenesis. 
The SN 1987A neutrino signal imposes the constraint
\begin{equation}
g_{\phi\gamma}  \lesssim 0.8 \times 10^{-8}\,{\rm GeV}^{-1},
\end{equation}
for $m_\phi\lesssim 200$~MeV~\cite{Caputo:2021rux}.  
However,  this bound can be avoided 
if axions strongly interact with nuclear matter via axion-Higgs mixing, effectively trapping them in the SN core.  
Even if the energy loss from the SN via axion emission 
falls below the bound,
additional constraints arise when
the axion predominantly decay into SM particles.
Let us summarize such constraints. 
First,  if axions escape freely from the SN core,  the SN 1987A $\gamma$-ray signal \cite{Caputo:2021rux,Caputo:2022mah,Hoof:2022xbe,Muller:2023vjm} and 
the diffuse cosmic $\gamma$-ray background place stringent limits on the axion coupling 
to photons~\cite{Caputo:2021rux,Caputo:2022mah}.
Additionally, axions produced in the protoneutron star could decay into photons, forming a fireball, which is a plasma of charged particles and photons~\cite{Diamond:2023scc}. 
The absence of a fireball signal in the energy range $0.2$ and $2\MeV$ from the Pioneer Venus Orbiter (PVO) satellite provides further constraints on the axion-photon coupling.
Second, axions should not deposit more energy in the SN mantle than is consistent with observations~\cite{Caputo:2022mah}.
Furthermore, 
a recent study has revealed that axions can be copiously produced in neutron star mergers, with their decay leading to fireball formation~\cite{Diamond:2023cto}. 
The non-observation of such a fireball following the emission of GW170817~\cite{LIGOScientific:2017vwq} 
places additional constraints on the axion-photon coupling. 
If the axion lifetime is sufficiently long to prevent fireball formation, the resulting $\gamma$-rays would still be constrained by Fermi-LAT observations~\cite{Dev:2023hax}.

We note that the aforementioned astrophysical constraints that arise when the axion dominantly decays to SM particles
can be alleviated in the presence of light dark particles  
that couple sufficiently strongly to the axion, 
while remaining weak enough to avoid disrupting the trapping of axions within the SN core.
This possibility  will be discussed in detail in subsequent sections. 
The BBN constraints apply for values of  $g_{\phi\gamma}$ smaller than about 
$3\times 10^{-9}\,{\rm GeV}^{-1}$ when the axion is heavier than $1$~MeV \cite{Cadamuro:2011fd}.   
However,  the disruption of BBN can be evaded if the axion lifetime is sufficiently short, 
which is achievable through axion-Higgs mixing and the presence of light dark particles 
coupled to the axion.
Lastly,  for $g_{\phi\gamma}$ larger than $2\times 10^{-7}\,{\rm GeV}^{-1}$,  
beam dump experiments  impose limits,  depending on the axion mass \cite{AxionLimits}.

Fig.~\ref{fig:bound-on-axion-to-photons} illustrates the allowed values of the axion-photon 
coupling as a function of the axion mass.
The gray-shaded regions indicates areas where the constraints can be alleviated 
due to axion-Higgs mixing and the presence of a dark sector coupled to the axion, which will be discussed later.
The light brown band represents the relationship, 
$g_{\phi\gamma} = \kappa_{\phi\gamma} m_\phi/m^2_h$, 
with the coefficient given by
\begin{equation}
\kappa_{\phi\gamma} = \frac{c_\gamma \alpha}{2\pi \kappa_1},
\end{equation}
which spans
from $10^{-4}$ to $10^{-2}$ for $c_\gamma/\kappa_1$  
values in the range of $0.1$ to $10$.

We conclude that the axion can successfully account for the observed baryon asymmetry 
of the universe while evading the experimental,  astrophysical,  and cosmological 
constraints on both axion-Higgs mixing and axion-photon coupling when  
\begin{equation}
    10^5\,{\rm GeV} \lesssim f_\phi \lesssim 10^7\,{\rm GeV},
\end{equation}
which corresponds to an axion mass in the range 
\begin{equation}
    {\cal O}(1)\,{\rm MeV} \lesssim m_\phi \lesssim {\cal O}(100)\,{\rm MeV},
\end{equation}
as determined by the relation $m_\phi =\kappa_1 m^2_h/f_\phi$.
In \SEC{sec:invisible}, we discuss the relaxation of the astrophysical and cosmological bounds shown as the gray shaded regions in 
\FIG{fig:bound-on-axion-to-photons}.
It
is worth noting that a portion of the viable parameter space will be accessible 
to future experimental searches such as SHiP~\cite{Alekhin:2015byh}.

\subsection{Axion cosmology}

We investigate the thermal population of axions in the early universe, 
focusing on the axion decay constant $f_\phi$ in the range of $10^5$ to $10^7$~GeV. 
At high temperatures,  prior to the electroweak phase transition, 
the axion coupling to the electroweak gauge sector ensures that axions remain in thermal equilibrium
with the SM thermal bath.
This equilibrium is  facilitated by the scattering process $f\bar f^\prime \leftrightarrow W  \phi$,
which occurs at a rate approximately given by $\Gamma  \sim \alpha^3_2 T^3/f^2_\phi$,
where $W$ denotes the SU$(2)$ gauge bosons,  and $f$ and $f^\prime$ are SM fermions. 
On the other hand,  the process $HH\leftrightarrow \phi\phi$ 
is subdominant,  with its rate approximately given by $\Gamma \sim (M/f_\phi)^4 T$.

At temperatures below the electroweak phase transition, 
Primakoff processes $f^\pm \gamma\to f^\pm \phi$,  mediated by the photon, 
continue to enable the production of axions from thermal bath.  
The Primakoff rate can be expressed as
\begin{equation}
\Gamma_{\rm P}  = c_1 \frac{\alpha^3}{f^2_\phi} T^3,
\end{equation}
at temperatures above $m_\phi$, 
where $c_1$ is  a constant of order unity.
This process decouples when its rate falls below the Hubble expansion rate. 
For $f_\phi$ above $10^5$~GeV,   decoupling 
occurs at temperatures above $m_\phi$,  approximately given by
\begin{equation}
    T_{\rm P} \sim 1\,{\rm GeV} \left( \frac{f_\phi}{10^6\,{\rm GeV}} \right)^2,
\end{equation} 
using the relation $m_\phi \sim m^2_h/f_\phi$,
and  a more precise estimation of $T_{\rm P}$ is available in Ref.~\cite{Bolz:2000fu}.

Another relevant mechanism contributing to axion production is Compton scattering, 
$f^\pm \gamma\to f^\pm \phi$,
mediated by SM fermions.
This process is induced by axion-Higgs mixing after the electroweak phase transition. 
For temperatures above $m_\phi$,  the Compton scattering rate is approximately given by 
 \begin{equation}
 \label{Compton-axion}
 \Gamma_{\rm C}
=  \sum_f c_2 \kappa^2_2  \alpha \left(\frac{m_f}{f_\phi}\right)^2 T,
 \end{equation}
where $c_2$ is a constant of order unity,  and $m_f$ denotes the mass of the fermion
involved in the scattering.   
The Compton scattering processes continue to maintain 
the axion in thermal equilibrium until the temperature drops to $m_\phi$
or below,  
provided that $f_\phi$ is less than approximately
$(\kappa^2_2/\kappa_1)\times 10^6$~GeV.
In deriving this result,   we use the fact that the Compton  rate
is approximated as $\Gamma_{\rm C} \approx c_2 \kappa^2_2 \alpha n_f/f^2_\phi$ 
for temperatures below $m_f$,  with $n_f$ being the fermion density.

Next,  we examine the production of axions at temperatures below $m_\phi$. 
Axion-Higgs mixing also facilitates the inverse decay process,  $e^+e^-\to \phi$,
which  becomes active at low temperatures,  with the decay rate scaling as $1/T$ 
for temperatures above $m_\phi$.
Using the mixing angle relation $\theta_{\rm mix} \simeq \kappa_2 m_h/f_\phi$, 
the inverse decay rate is expressed by
\begin{equation}
\Gamma_{\rm D} = \frac{c_3 \kappa^2_2}{8\pi} \left( \frac{m_f}{f_\phi} \right)^2 \frac{m^2_\phi}{T},
\end{equation}
at temperatures  above $m_\phi$,  where $c_3$ is  a constant of order unity.
It is evident that the inverse decay  becomes more efficient than 
the Compton scattering, 
 $e^\pm \gamma \to e^\pm \phi$,  at $T$ below about $2m_\phi$.
Using the relation $m_\phi = \kappa_1 m^2_h/f_\phi$,
the ratio of the rate for $e^+e^-\to \phi$ 
to the Hubble expansion rate is approximately  
\begin{equation}
\label{inverse-ep-to-axion}
\frac{\Gamma_{\rm D}}{H} 
\sim  
\frac{\kappa^2_2}{\kappa_1}
\left( \frac{g_\ast}{10} \right)^{-1/2}
\left( \frac{f_\phi}{10^6\,{\rm GeV}} \right)^{-1},
\end{equation}
at temperatures around $m_\phi$,
where  $g_\ast(T)$ denotes the number of relativistic degrees of freedom. 
This indicates that the inverse decay process becomes efficient at or before 
the temperature reaches $m_\phi$,  if $f_\phi$ does not exceed approximately
$(\kappa^2_2/\kappa_1)\times 10^6$~GeV.
Meanwhile, 
as highly suppressed compared to $e^+e^-\to \phi$,   the inverse decay process 
$\gamma\gamma \to \phi$  is always subdominant in the production of axions.

As a consequence of the coupling to both the Higgs and the electroweak gauge sectors,   
axions remain in thermal equilibrium with the SM plasma until they become 
non-relativistic,  provided the axion decay constant falls within the range
\begin{equation}
\label{axion-thermalization}
 10^5\,{\rm GeV}
  \lesssim f_\phi \lesssim  
 \frac{\kappa^2_2}{\kappa_1}  \times 
  10^6\,{\rm GeV}.
\end{equation}
In this regime,  thermal equilibrium leads to a Boltzmann suppression of the axion abundance.
Throughout this paper,  we assume this scenario,  with appropriate values 
for $\kappa_1$ and $\kappa_2$.
For values of $f_\phi$ exceeding the upper bound in the above range, 
Compton processes freeze out at temperatures above $m_\phi$, 
resulting in axion decoupling from thermal bath while still relativistic. 
However,  re-thermalization may occur if the inverse decay process $e^+e^- \to \phi$ 
becomes sufficiently efficient at lower temperatures.\footnote{
While this work explores a minimal scenario,  one can also incorporate an axion coupling 
to the QCD anomaly to enhance its interactions with the SM thermal bath.
In this case,   even for $f_\phi$ much above $10^7$~GeV,
the axion remains in thermal equilibrium via scattering processes 
 $gg\leftrightarrow g \phi$ and $qq \leftrightarrow g \phi$ 
until the temperature approaches the QCD confinement scale $\Lambda_{\rm QCD}$.
Here $g$ and $q$ denote the gluons and quarks,  respectively.
Below $\Lambda_{\rm QCD}$,  pion scattering processes $\pi\pi \leftrightarrow \pi \phi$ 
can effectively thermalize the axion until the temperature decreases to a few tens of MeV 
for $f_\phi$ below $10^7$~GeV.  
}

\section{Dark photons
\label{sec:dark-photon}}

Light dark particles coupled to the axion can alleviate the constraints 
on the axion-photon coupling
imposed by BBN and SN explosions by enhancing the axion decay rate 
while simultaneously suppressing the branching ratio of axion decay into 
SM particles.
Since axion interactions are governed by the global U$(1)_\phi$ symmetry,  
dark photons,  whose coupling to the axion naturally arises from the
U$(1)_\phi$ anomaly, 
emerge as a compelling candidate for these light dark particles.\footnote{
Although we do not consider this possibility in detail here, 
an interesting alternative 
involves the inclusion of dark fermions coupled to the axion. 
Their interactions are constrained by U$(1)_\phi$,  and are generally of two types, given by
\begin{equation}
    \frac{k_1 \partial_\mu \phi}{f_\phi} \bar f \gamma^\mu \gamma_5 f
    + m_f e^{i k_2 \frac{\phi}{f_\phi} } \bar f f,
    \nonumber 
\end{equation}
where $m_f$ denotes the mass of $f$, and $k_1$ and $k_2$ are constants of order unity. 
Physical quantities depend on a combination of $k_1$ and $k_2$ that remains invariant under chiral field rotations.
These interactions give rise to 
$y_f \phi \bar f f$, with a Yukawa 
coupling given by $y_f \sim m_f/f_\phi$, 
indicating that the axion couplings to dark fermions are more constrained compared to those with dark photons.
} 
Axion-coupled dark photons 
can also play an important role in cosmology~\cite{Agrawal:2017eqm,Kitajima:2017peg,Agrawal:2018vin,Kitajima:2021bjq,Nakagawa:2022knn},  
with their impact depending on their masses generated through either the Higgs or Stueckelberg mechanism~\cite{Stueckelberg:1938hvi}.
Although not considered in this work,  
dark photons may exhibit kinetic mixing with ordinary photons. 
This kinetic mixing,  which should be sufficiently small to remain consistent with 
various constraints~\cite{Fabbrichesi:2020wbt,Caputo:2021eaa},  
would allow dark photons to interact with SM particles.

We consider a dark photon,  $\gamma^\prime$,
to which the axion has an anomalous coupling given by
 \begin{equation}
    \frac{g_{\phi\gamma^\prime}}{4} \phi F^\prime_{\mu\nu} \tilde F^{\prime\mu\nu},
\end{equation}
with its mass $m_{\gamma^\prime}<m_\phi/2$.
The axion coupling to dark photons can be written,
$g_{\phi\gamma^\prime} = c_{\gamma^\prime}\alpha^\prime/(2\pi f_\phi)$,
where $\alpha^\prime$ is the dark gauge coupling,  and $c_{\gamma^\prime}$ denotes
the anomaly coefficient,  generally  a rational number of order unity.   
We assume that any dark fermions coupled to the dark photon,  if they exist, 
are all heavier than the reheating temperature of the universe,  $T_{\rm reh}$, 
after primordial inflation.
Consequently,   the cosmological properties of the dark photon are governed by the axion.

For the axion decay constant within the range given in Eq.~(\ref{axion-thermalization}),   
axions remain in thermal equilibrium until the temperature drops to $m_\phi$
or below.  
Thus,  in the absence of dark fermions, 
dark photons can be produced from thermal bath through 
the scattering process $\phi\phi \to \gamma^\prime \gamma^\prime$ 
and the decay process $\phi\to \gamma^\prime\gamma^\prime$.
The scattering rate is estimated by
\begin{equation}
    \Gamma_{\phi\phi\to \gamma^\prime\gamma^\prime}
    = c^\prime_1 g^4_{\phi\gamma^\prime} T^5,
\end{equation}
at temperatures above $m_\phi$,
where $c^\prime_1$ is a constant of order unity. 
Due to the temperature dependence of its rate, this process becomes 
efficient only at high temperatures, 
specifically above the decoupling temperature of the dark photon, 
approximately given by 
\begin{equation}
\label{Tdec-dark-photon}
    T_{\rm dec} \approx 
    5\times 10^4\,{\rm GeV}
    \left(\frac{g_{\phi\gamma^\prime}}{10^{-8}\,{\rm GeV}^{-1}}\right)^{-4/3},
\end{equation}
taking $c^\prime_1 \approx 1$  and $g_\ast \approx 100$. 
At temperatures below $T_{\rm dec}$,   the dynamics of axion decays crucially depends on 
the ratio of the decay rate to the Hubble expansion rate
at temperatures around $m_\phi$,  which is given by
\begin{equation}
\frac{\Gamma_{\phi\to\gamma^\prime\gamma^\prime}}{H} \approx
10^{-2} 
 \left(
\frac{g_{\phi\gamma^\prime}}{10^{-8}\,{\rm GeV}^{-1}}
    \right)^2
    \left( \frac{m_\phi}{10\,{\rm MeV}} \right),
\end{equation}
for a dark photon much lighter than the axion, 
where we have taken $g_\ast\approx 10$.  
If this ratio exceeds unity,  dark photons thermalize via the process 
$\phi \leftrightarrow \gamma^\prime \gamma^\prime$,
even after the scattering process $\phi\phi \leftrightarrow \gamma^\prime \gamma^\prime$ 
freezes out,
as the axion remains in equilibrium until the temperature drops to or below $m_\phi$.

In this work,  we explore three distinct classes of 
axion-coupled dark photons:
$(i)$ unstable dark photons,  into which the axion predominantly decays,  thereby suppressing 
its decay into SM particles, 
$(ii)$ stable dark photons with masses exceeding $10$~keV,  contributing to cold dark matter,  and
$(iii)$ ultralight dark photons with masses below the eV scale,  contributing to dark radiation.
The abundance of these dark photons depends on the strength of their coupling to the axion and 
whether they ever reached thermal equilibrium.

\subsection{Invisible axion decay
\label{sec:invisible}}

We introduce a massive dark photon,  denoted as  $\gamma^\prime_{\rm M}$, 
that is coupled to the axion to relax the constraints on the axion-photon coupling that arise from the SN 1987A  $\gamma$-ray signal,  
the diffuse cosmic $\gamma$-ray background,  and observations of low-energy supernovae. 
These astrophysical limits are applicable for $m_\phi$ below about $200$~MeV,
as the relation $m_\phi f_\phi = \kappa_1 m^2_h$ holds for the axion mass and decay constant.
Considering this,  we focus in this subsection on the axion mass within the range
\begin{equation}
3\,{\rm MeV} < m_\phi  < 200\,{\rm MeV},
\end{equation}
where the lower bound ensures that axions in thermal bath become  
non-relativistic before BBN. 
In this mass range,  the axion decays to photons and electrons
among the SM particles,  
with the relative decay rates described by   
\begin{equation}
\frac{\Gamma_{\phi \to \gamma\gamma}}{\Gamma_{\phi\to e^+e^-}}
\sim 
\frac{10^{-5}}{\kappa^2_2} 
\left(\frac{m_\phi}{10\,{\rm MeV}} \right)^2,
\end{equation}
where the decay to electrons is induced by axion-Higgs mixing.
As a result, this mixing reduces the axion decay length, thereby alleviating constraints from 
$\gamma$-ray burst observations.

The constraints on the axion-photon coupling from SN 
explosions and the neutron merger for GW170817 are 
sensitive to the branching ratio of axion decay into SM particles.  
To suppress the branching ratio,   
the axion coupling to dark photons should be 
much larger than the coupling to electrons, leading to the condition 
\begin{equation} 
\label{inv-axion-decay-condition}
g_{\phi\gamma^\prime_{\rm M}} >  
 \frac{2\kappa_2  m_e}{m_\phi f_\phi}
\simeq 
\left(\frac{\kappa_2}{\kappa_1}\right) \times 
10^{-7}\,{\rm GeV}^{-1},
\end{equation}
where the latter equality follows from $m_\phi f_\phi = \kappa_1 m^2_h$,
rendering the lower bound on $g_{\phi\gamma^\prime_{\rm M}}$ insensitive to
$m_\phi$ and $f_\phi$.  
The perturbativity of gauge interactions 
requires that $g_{\phi\gamma^\prime_M}$
remain below approximately $1/f_\phi$, unless the associated anomaly 
coefficient is significantly greater than unity.
This implies that the branching ratio of axion decay into SM particles can be suppressed up to 
\begin{equation}
    {\rm Br}(\phi\to {\rm SM})
    \gtrsim
    10^{-4}
    \left(\frac{\kappa_2}{0.1}\right)^2
    \left(\frac{m_\phi}{10\,{\rm MeV}}\right)^{-2},
\end{equation}
relaxing the constraints from SN explosions and the neutron merger for GW170817.
The branching ratio can be further suppressed by introducing additional dark photons. 
It is important to note that the axion coupling to a dark sector
is constrained by the requirement that axions remain trapped within the SN core 
through their interactions with nuclear matter.  
This requires that the axion decay length should be much longer than 
the axion mean free path in the SN core,
ensuring that axions are absorbed through the inverse pion conversion 
and inverse bremsstrahlung before they decay to dark particles~\cite{Turner:1987by,Caputo:2024oqc}.
For the coupling constant given by
\begin{equation}
\label{axion-DP-upper}
 g_{\phi\gamma^\prime_{\rm M} } <
10^{-6}\,{\rm GeV}^{-1}
  \left(  \frac{m_\phi}{10\,{\rm MeV}} \right)^{-3/2},
\end{equation}
the axion decay length at rest is larger than $\mathcal{O(}10)$~km,
the size of the core,  
thereby not interfering with the trapping processes.
We assume the existence of dark photons that couple to the axion, 
with coupling strengths constrained by (\ref{inv-axion-decay-condition})
and (\ref{axion-DP-upper}).

Axion-coupled dark photons can also alleviate the constraints from BBN 
by allowing the axion to decay sufficiently early, 
before the onset of BBN. 
For axion masses in the range of MeV to a few hundred MeV,
the BBN constraints can be evaded if the axion lifetime is shorter 
than approximately  
$10^2\,{\rm sec} / (m_\phi/{\rm MeV})$~\cite{Cadamuro:2011fd}.
This condition is met 
if dark photons couple to the axion with 
\begin{equation}
    g_{\phi\gamma^\prime_M} >
    4\times 10^{-9}\,{\rm GeV}^{-1}
    \left( \frac{m_\phi}{10\,{\rm MeV}} \right)^{-1},
\end{equation} 
or if the axion-Higgs mixing, which induces axion decay to electrons, satisfies
\begin{equation}
\sin^2 \theta_{\rm mix} > 10^{-11}.
\end{equation}
In our model, the axion-Higgs mixing is already sufficiently large to meet this requirement,
except in the case where $\kappa_2$ is extremely small.
Dark photons with a coupling given by 
Eq.~(\ref{inv-axion-decay-condition}) further reduce the axion lifetime.
This indicates that the primary role of dark photons is to alleviate the astrophysical constraints arising from SN explosions.

If the dark photon couples to the axion more strongly than the electron, dark photons undergo thermalization at temperatures above $m_\phi$ through forward
and inverse decay processes,  as indicated by the fact that the process $e^+e^-\to\phi$
becomes efficient before the temperatures reaches $m_\phi$.    
This implies that dark photons should remain in thermal equilibrium until they become 
highly non-relativistic,  ensuring their number density is sufficiently Boltzmann suppressed.
Failure to meet this condition would result in dark photons overclosing the universe
if they are stable,  or contributing excessive dark radiation if they are very light.  
However,  a challenge arises in achieving a strong Boltzmann suppression of 
the dark photon density, 
because the axion decouples from the SM thermal bath in the non-relativistic regime
and serves as the only mediator connecting the dark photon to the SM sector.\footnote{
In the absence of effective Boltzmann suppression,  one alternative is to rely on late-time entropy production to dilute the dark photon density. 
However, this approach also dilutes the baryon asymmetry,  
necessitating an additional mechanism to generate baryon asymmetry.
}

To suppress the axion decay into SM particles while avoiding the cosmological difficulties,   
we investigate a scenario in which the dark photon,   
into which the axion predominantly decays,
is not stable and decays to ultralight dark fermions shortly after the dark photon freezes out. 
We assume that these dark fermions have a mass below the eV scale, 
and does not couple to either the axion or the SM sector. 
Under these conditions,  
dark fermions produced from the out-of-equilibrium decays 
of dark photons can contribute to dark radiation in an amount consistent 
with cosmological constraints,  even if the Boltzmann suppression of the dark photon density
is not significantly strong. 
We will now examine the conditions that render this scenario viable.

Both the axion and dark photon remain in thermal equilibrium at temperatures above $m_\phi$. 
At lower temperatures,   the dark photon undergoes freeze-out  when the inverse decay rate
for the process $\gamma^\prime_{\rm M} \gamma^\prime_{\rm M} \to \phi$,
\begin{equation}
\Gamma_{\rm ID}
\approx
\frac{n_\phi}{n_{\gamma^\prime_{\rm M}}} \frac{1}{64\pi} g^2_{\phi\gamma^\prime_{\rm M}} m^3_\phi,
\end{equation}
falls below the Hubble expansion rate.
Here,   $n_i$ is the number density of the species $i$, 
and we take the dark photon mass to be $3\,{\rm MeV} < m_{\gamma^\prime_{\rm M}} < m_\phi/2$,
with the lower bound ensuring that the dark photon becomes non-relativistic 
prior to BBN. 
After freeze-out,  dark photons dominantly decay into ultralight dark fermions.
The contribution of these dark fermions to the effective number of neutrino species
can be estimated as
\begin{equation}
\Delta N_{\rm eff} 
\approx \left(\frac{T_{\rm fo}}{T_{\rm decay}} \right)
\left( \frac{m_{\gamma^\prime_{\rm M}}}{T_{\rm fo}} \right)^{5/2}
e^{-m_{\gamma^\prime_{\rm M}}/T_{\rm fo}},
\end{equation}
under the assumption that $T_{\rm decay} < T_{\rm fo}  \lesssim m_{\gamma^\prime_{\rm M}}/3$,
where $T_{\rm fo}$ and  $T_{\rm decay}$ denote the freeze-out and decay
temperature of the dark photon,  respectively.   
Although the out-of-equilibrium decay of axions also produces dark fermions,   
their contribution is negligible due to the large Boltzmann suppression of 
the axion density compared to that of the dark photon. 
The freeze-out temperature $T_{\rm fo}$ is determined by the condition
 $\Gamma_{\rm ID} \approx H$ at $T=T_{\rm fo}$,
which leads to 
\beq
\frac{m_\phi-m_{\gamma^\prime_{\rm M}}}{T_{\rm fo}} 
&\approx&
11 +  2 \ln\left(\frac{m_\phi/T_{\rm fo}}{10} \right)
\nonumber \\
&&
-\, \frac{3}{2} \ln\left(\frac{m_{\gamma^\prime_{\rm M}}/m_\phi}{0.3} \right)
+  \ln\left( \frac{m_\phi}{10\,{\rm MeV} } \right)
\nonumber \\
&&
+\, 2 \ln\left( \frac{g_{\phi\gamma^\prime_{\rm M}}}{10^{-6}\,{\rm GeV}^{-1} } \right),
\eeq
under the conditions,   $3\,{\rm MeV} < m_{\gamma^\prime_{\rm M}} < m_\phi/2$ and 
$T_{\rm decay} < T_{\rm fo}  \lesssim m_{\gamma^\prime_{\rm M}}/3$.
For the values of  $f_\phi$ and $m_\phi$ considered,    
the freeze-out temperature lies approximately in the range $m_\phi/5$ to $m_\phi/15$,
where we have used that 
the perturbative bound on the dark gauge coupling imposes
$g_{\phi\gamma^\prime_{\rm M}}\lesssim 1/f_\phi$ unless the associated anomaly coefficient
is significantly larger than unity. 
As a result, $\Delta N_{\rm eff}$ is at most of order $0.1$,  which is consistent with current 
cosmological constraints. 
Furthermore,  it is important to note that the Boltzmann suppression of the dark photon density, 
$e^{-m_{\gamma^\prime_{\rm M}}/T_{\rm fo}}$,  is not substantial for $T_{\rm fo}$ 
within the indicated range. 
This explains why the dark photon should not be a stable particle.

A final remark concerns the decay of the dark photon to ultralight dark fermions. 
The discussion thus far assumes that the dark fermions have a mass below the eV scale,  and decouple from thermal bath before the electroweak phase transition if they reached thermal equilibrium in the early universe. 
This decoupling is necessary to ensure that their contribution to $\Delta N_{\rm eff}$ remains
on the order of $0.1$ or less. 
The decay of the dark photon can proceed via the interaction 
\begin{equation}
{\cal L} = \varepsilon  A^\prime_\mu \bar \psi \gamma^\mu \psi, 
\end{equation}
with $\varepsilon \ll 1$.
The decay temperature of the dark photon is lower than its freeze-out temperature if 
\begin{equation}
 \varepsilon < 10^{-9} 
 \left(\frac{m_{\gamma^\prime_{\rm M}}}{10\,{\rm MeV}} \right)^{1/2},
\end{equation} 
for $T_{\rm fo} \lesssim m_{\gamma^\prime_{\rm M}}/3$.   
If the dark fermion $\psi$ is charged under a hidden U$(1)$ gauge symmetry, 
a small value of $\varepsilon$ can naturally arise from kinetic mixing between the dark photon $\gamma^\prime_{\rm M}$ and the hidden U$(1)$ gauge boson. 
In this scenario,  the hidden U$(1)$ gauge boson is assumed to be much heavier than 
the electroweak scale, 
ensuring its decoupling from thermal bath at high temperatures.

\subsection{Dark matter}

Dark photons are a compelling candidate for dark matter,
with several mechanisms proposed for generating 
the correct relic abundance,
such as freeze-out/freeze-in 
processes~\cite{Redondo:2008ec,Pospelov:2008jk,Chu:2011be,Fradette:2014sza,Essig:2015cda,Berger:2016vxi}, gravitational production~\cite{Graham:2015rva,Ema:2019yrd}, coherent oscillations~\cite{Nakayama:2019rhg,Kitajima:2023fun}, and nonthermal production through axion couplings~\cite{Agrawal:2018vin,Co:2018lka,Bastero-Gil:2018uel}.
In this work, we focus on the freeze-in production of dark photon dark matter.
Previous studies~\cite{Redondo:2008ec,Pospelov:2008jk,Chu:2011be,Fradette:2014sza,Essig:2015cda,Berger:2016vxi}
have considered freeze-in production via kinetic mixing with the SM photon, 
which is severely constrained by experimental limits~\cite{Fabbrichesi:2020wbt,Caputo:2021eaa}.
Here, we propose an alternative scenario where 
the axion associated with electroweak baryogenesis assists the freeze-in production of dark photons, 
assuming that the kinetic mixing is sufficiently suppressed.

To account for the observed dark matter density
and the formation of structure in the universe,   
we introduce a stable dark photon, 
$\gamma^\prime_{\rm DM}$,  with mass $m_{\gamma^\prime_{\rm DM}} > 10$~keV but
below $m_\phi/2$.
As axions remain in thermal equilibrium until they become non-relativistic, these dark photons can be produced from the thermal bath.
However, they would be overproduced if their freeze-out occurs while still relativistic.
To prevent the overproduction, the scattering process 
$\phi \phi \to \gamma^\prime_{\rm DM} \gamma^\prime_{\rm DM}$ 
must be strongly suppressed after reheating of the inflation. 
This imposes an upper bound on the reheating temperature as
\beq
    T_{\rm reh} <  10^6 \,{\rm GeV} \left( 
    \frac{g_{\phi\gamma^\prime_{\rm DM}}}{10^{-9} \,{\rm GeV}^{-1}} 
    \right)^{-4/3},
\eeq
where $g_{\phi\gamma^\prime_{\rm DM}}$ represents the axion coupling to 
$\gamma^\prime_{\rm DM}$, 
and we have used the relation (\ref{Tdec-dark-photon}).
An additional constraint arises from the potential thermalization via the process 
$\phi \to \gamma^\prime_{\rm DM} \gamma^\prime_{\rm DM}$ at low temperatures.
This can be avoided if the following condition is satisfied  
\begin{equation}
\label{Treheating-condition}
    g_{\phi\gamma^\prime_{\rm DM}} <
 10^{-8} \,{\rm GeV}^{-1}
    \left(
\frac{m_\phi}{10\,{\rm MeV}}
    \right)^{-1/2},
\end{equation}
which follows from the requirement that the decay rate of 
$\phi \to \gamma^\prime_{\rm DM} \gamma^\prime_{\rm DM}$ should remain lower 
than the Hubble expansion rate until  axions decouple from thermal bath,  occurring 
at a temperature around $m_\phi/10$ or below. 
The dark photon density produced from the out-of-equilibrium decay of 
axions is negligibly small,  because  it is Boltzmann suppressed by a factor of
 $e^{-m_\phi/T}$ with $T\approx m_\phi/10$, 
 and further reduced by the branching fraction
${\rm Br}(\phi\to\gamma^\prime_{\rm DM}\gamma^\prime_{\rm DM})$.
The branching ratio is highly suppressed because the axion predominantly decays 
into unstable dark photons,  which couple more strongly to the axion than electrons.
This also indicates that non-thermal production of dark photons during 
axion oscillations,  which occur during the electroweak phase transition,
is not efficient.

\begin{figure}[t!]
\centering
\includegraphics[width=8.5cm]{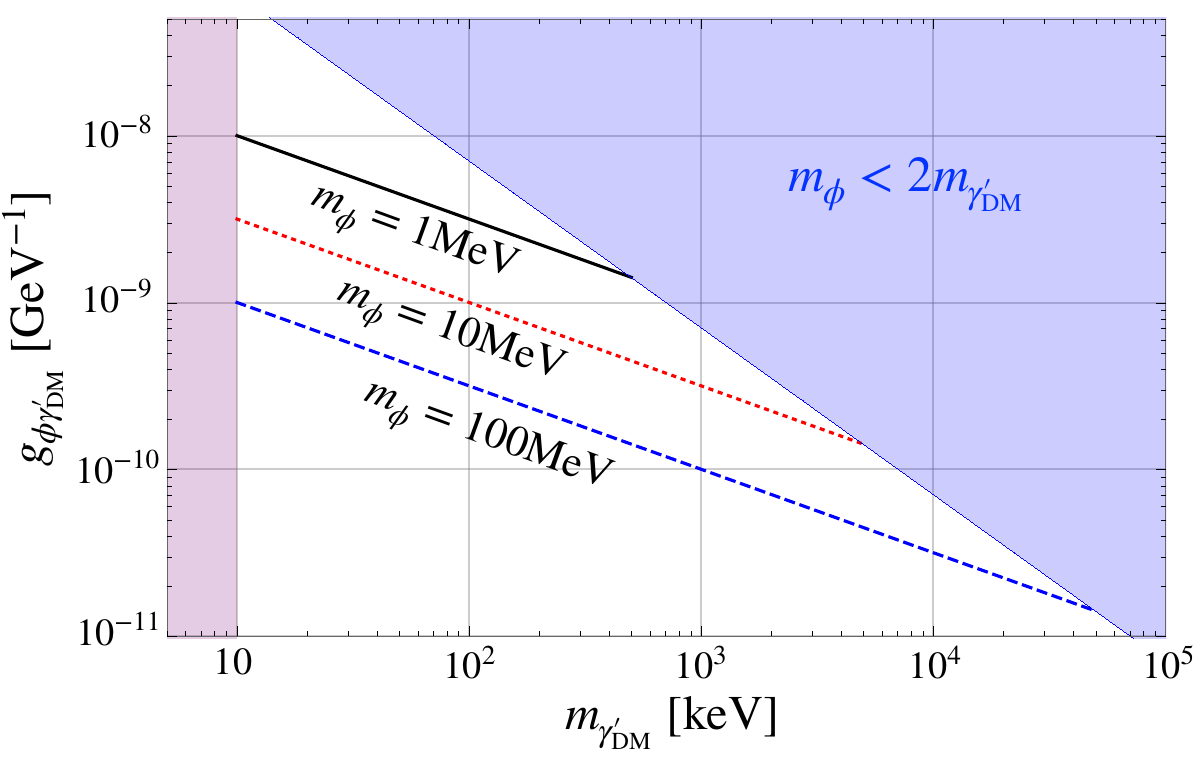}
\caption{
Dark photon relic density as a function of its mass and coupling to the axion. 
Dark photons produced via the freeze-in mechanism account for the observed 
dark matter density along the curve for a given value of the axion mass.  
The freeze-in process occurs through the decay of axions in thermal bath, 
thereby excluding the blue-shaded region.
The lower bound on the dark photon mass comes from constraints 
on structure formation in the universe.
The observed baryon asymmetry is generated through axion-induced 
electroweak baryogenesis for axion masses ranging from the MeV to GeV scales.
}
\label{fig:abundance}
\end{figure}

Although dark photons never reach thermal equilibrium,  they can still be produced via the freeze-in mechanism~\cite{McDonald:2001vt,Choi:2005vq,Kusenko:2006rh,Petraki:2007gq,Hall:2009bx} through the decay of axions in thermal bath.
The resulting dark photon density can be approximately estimated as~\cite{Bernal:2017kxu}:
\beq
    \Omega_{\gamma'_{\rm DM}} &\approx&
\Omega_{\rm DM}
\left( \frac{m_{\gamma^\prime_{\rm DM}}}{100\,{\rm keV}} \right)
\left( \frac{g_{\phi\gamma^\prime_{\rm DM}}}{10^{-10}\,{\rm GeV}^{-1}} \right)^2
\nonumber \\
&&
\times
\left(\frac{g_\ast}{10}\right)^{-3/2}
\left( \frac{m_\phi}{10\,{\rm MeV}} \right), 
\label{abundance}
\eeq
under the assumption that the initial dark photon abundance is negligibly small.
Here we denote the observed density parameter for dark matter as $\Omega_{\rm DM}$, and $g_\ast(T)$ is taken to be the value at a temperature around 
 $m_\phi$,  the point at which the freeze-in process is most efficient.
This expression indicates that 
massive dark photons can constitute the dominant component of dark matter 
in the universe across a wide range of the parameter space. 
We have confirmed that the approximate relation above is consistent with the dark photon
relic density obtained from numerically solving the Boltzmann equation.
Furthermore,  we note that the condition (\ref{Treheating-condition}), 
necessary to prevent the thermalization of dark photons, 
is always satisfied within the parameter space where the dark photon density does 
not exceed the observed dark matter density.
In addition, the dark photon dark matter can be produced from the quantum fluctuation during inflation \cite{Graham:2015rva,Ema:2019yrd}.
The abundance is sufficiently suppressed  
if the Hubble parameter during inflation is below approximately
$10^9$~GeV.

Fig.~\ref{fig:abundance} depicts  the parameter space in which the dark photon, 
interacting with the SM sector solely through the axion, 
accounts for the observed dark matter density.
The black solid,  red dotted,  and blue dashed lines represent the contours corresponding to
 $\Omega_{\gamma'_{\rm DM}} \simeq \Omega_{\rm DM}$ 
for the respective values of the axion mass. 
Successful electroweak baryogenesis constrains
$m_\phi$ to the range of approximately MeV to GeV scales. 
Dark photons are not produced in the blue shaded region, 
as the decay of axion to dark photons is kinematically forbidden. 
The purple shaded region,  
corresponding to dark photon masses below $10$~keV,  
is excluded because 
dark photons produced through the freeze-in mechanism do not facilitate structure formation
in agreement with observational constraints~\cite{DEramo:2020gpr}.

\subsection{Dark radiation}

An ultralight dark photon,  denoted as $\gamma^\prime_{\rm DR}$, 
with mass $m_{\gamma^\prime_{\rm DR}} \lesssim 1$~eV and coupling 
to the axion represented by $g_{\phi\gamma^\prime_{\rm DR}}$, 
and behaves as dark radiation because dark photons are produced  
from thermal bath 
through the processes $\phi\phi\to \gamma^\prime_{\rm DR}\gamma^\prime_{\rm DR}$ and 
$\phi\to \gamma^\prime_{\rm DR}\gamma^\prime_{\rm DR}$.  
Dark photons get thermalized via the scattering process 
$\phi\phi\leftrightarrow \gamma^\prime_{\rm DR}\gamma^\prime_{\rm DR}$  
if its decoupling temperature $T_{\rm dec\, {\rm DR}}$ is lower than $T_{\rm reh}$.
It should be noted that the condition $T_{\rm dec\,{\rm DR}} < T_{\rm reh}$ 
is only applicable for dark photons with mass below the eV scale.   
Otherwise,  freeze-out dark photons would result in overclosure of the universe.
Upon decoupling from  thermal bath, dark photons contribute 
to the effective number of neutrino species as follows
\begin{equation}
\Delta N_{\rm eff} 
     = \frac{4g}{7}\left( \frac{11}{4} \right)^{4/3}
    \left(\frac{g_{\ast s}(T)}{g_{\ast s}(T_{\rm dec\, DR}) } \right)^{4/3},
\end{equation}
because the dark photon temperature is given by
\begin{equation}
    T_{\gamma^\prime_{\rm DR}}
    = \left(\frac{g_{\ast s}(T)}{g_{\ast s}(T_{\rm dec\,DR}) } \right)^{1/3} T,
\end{equation}
which derives from the entropy conservation 
and the fact that the dark photon temperature decreases according to
the relation $R(T) T_{\gamma^\prime_{\rm DR}} = R(T_{\rm dec\,DR}) T_{\rm dec\,DR}$
after decoupling,
with $R$ being the scale factor. 
Here, $g = 3$ represents the number of degrees of freedom associated 
with the dark photon, and $g_{\ast s}$ denotes the effective number of 
degrees of freedom contributing to the entropy.
For scenarios in which the dark photon decouples at a temperature above the electroweak phase transition, its contribution is approximately given by
\begin{equation}
    \Delta N_{\rm eff} \approx 0.07,
\end{equation}
in the present universe.
Consequently,  the existence of multiple ultralight dark photons with masses below the eV scale is consistent with the Planck constraints, 
$\Delta N_{\rm eff}\lesssim 0.3$~\cite{Planck:2018vyg},  
potentially leading to observable cosmological effects~\cite{Archidiacono:2013fha}.

\section{Conclusions
\label{sec:conclusion}}

If the SM is extended to include an axion that couples to both the Higgs and electroweak gauge sectors, 
the matter-antimatter asymmetry in the universe can be explained through electroweak 
baryogenesis. 
For this to occur,  the axion decay constant should lie between approximately $10^5$ and $10^7$~GeV,
corresponding to axion masses ranging from the MeV to GeV scales.
It is plausible that the axion responsible for baryogenesis interacts with dark photons via a natural coupling arising from the U$(1)$ anomaly. 
These axion-coupled dark photons can suppress the branching ratio of axion decay to SM particles,
thereby relaxing the constraints on the axion that arise from SN explosions and BBN.
Moreover,  axion-coupled dark photons can constitute the main component of dark matter 
as they are produced from 
the decay of axions in thermal bath through the freeze-in mechanism. 
Additionally,  ultralight axion-coupled dark photons may provide an observable contribution 
to dark radiation.
Unlike other scenarios for electroweak baryogenesis,  
our model can be probed through searches for axion-like particles rather than relying on 
EDM and LHC experiments.
This is because the axion interacts with the SM feebly, 
via couplings suppressed by a large decay constant.

\section*{Acknowledgments}
K.S.J.  expresses gratitude to Particle Theory Group at the University of Oxford for their generous hospitality during his sabbatical visit. 
K.S.J.  and J.H.K.  acknowledge support from the National Research Foundation (NRF) of Korea, funded by the Korean government under Grant No. RS-2023-00249330.

\appendix

\bibliography{reference}

\end{document}